\documentclass[fleqn,usenatbib,useAMS]{mnras}

\usepackage{newtxtext,newtxmath}

\usepackage[T1]{fontenc}
\usepackage{ae,aecompl}

\usepackage{graphicx}	
\usepackage{amsmath}	
\usepackage{amssymb}	
\usepackage{comment}
\usepackage[section]{placeins}

\title[Breaking the centrifugal barrier]{Breaking the centrifugal barrier to giant planet contraction by magnetic disc braking}

\author[S. Ginzburg and E. Chiang]{
Sivan Ginzburg$^{1}$\thanks{E-mail: ginzburg@berkeley.edu}\thanks{51 Pegasi b Fellow.}
and Eugene Chiang$^{1,2}$
\\
$^{1}$Department of Astronomy, University of California at Berkeley, CA 94720-3411, USA\\
$^{2}$Department of Earth and Planetary Science, University of California at Berkeley, CA 94720-4767, USA
}

\date{Accepted XXX. Received YYY; in original form ZZZ}

\pubyear{2019}

\begin{document}
\label{firstpage}
\pagerange{\pageref{firstpage}--\pageref{lastpage}}
\maketitle

\begin{abstract}
During the runaway phase of their formation, gas giants fill their
gravitational spheres of influence out to Bondi or Hill radii. When runaway ends, planets shrink several orders of magnitude in radius until they are comparable in size to present-day Jupiter; in 1D models, the contraction
occurs on the Kelvin--Helmholtz time-scale $t_{\rm KH}$, 
which is initially a few thousand years. 
However, if angular momentum is conserved, 
contraction cannot complete, as
planets are inevitably spun up to their breakup periods $P_{\rm break}$. 
We consider how a circumplanetary disc (CPD) can de-spin a primordially magnetized gas giant and remove the centrifugal barrier, provided the disc is hot enough to couple to the magnetic field, a condition that is easier to satisfy at later times.
By inferring the planet's magnetic field from its
convective cooling luminosity, we show that magnetic spin-down times are shorter than contraction times throughout post-runaway contraction: $t_{\rm mag}/t_{\rm KH}\sim(P_{\rm break}/t_{\rm KH})^{1/21}\lesssim 1$.
Planets can spin down until they corotate with the CPD's magnetospheric truncation radius, at a period $P_{\rm max}/P_{\rm break} \sim (t_{\rm KH}/P_{\rm break})^{1/7}$. By the time the disc disperses, $P_{\rm max}/P_{\rm break}\sim$ 20--30;
further contraction at fixed angular momentum can spin planets back up to 
$\sim$$10 P_{\rm break}$, potentially explaining  observed rotation periods of giant planets and brown dwarfs.
\end{abstract}




\section{Introduction}\label{sec:introuction}

Gas giants are thought to form when gas from an ambient circumstellar
disc cools atop a rocky or icy core
\citep[e.g.][]{BodenheimerPollack86,Pollack1996,Ikoma2000}. 
The planet's proto-atmosphere extends to the smaller of the Bondi radius and the Hill radius, both of which are several orders of magnitude larger than present-day Jupiter. Atmospheric gas cools and contracts on a Kelvin--Helmholtz (KH) time-scale, allowing fresh nebular gas to flow in and take its place.
Once the atmosphere outweighs the core, the KH time-scale 
begins to decrease with increasing mass,
and the planet cools and grows at an ever faster, `runaway' rate. 

Accretion is also limited by the nebula's ability to supply gas at a sufficient rate to the Bondi/Hill radius to keep up with the planet's increasingly shorter cooling and contraction time. 
The rate of supply is capped by the 
radial transport rate through the disc, 
and further limited by the opening of annular gaps
(depleted cavities) around the planet's orbit.
Both of these effects can put an end to runaway growth 
\citep{LinPapaloizou93,Kley99,TanigawaIkoma2007,Lissauer2009,TanigawaTanaka2016,GC_FinalMass,Lee2019}.
Post-runaway planetary accretion is not zero; it persists 
up to the eventual dispersal of the disc in a few million years' time
\citep{Mamajek2009,Pfalzner2014}, and can be responsible for doubling the planet's mass or more
\citep{Mordasini2012,GC_Endgame}.

In the post-runaway phase, because of insufficient gas supply, 
the planet no longer fills the
Bondi/Hill radius but detaches from the nebula
to contract on the KH time-scale 
\citep{Bodenheimer2000,Mordasini2012,Mordasini2017,GC_Endgame}.
Previous one-dimensional models assumed the planet is free to shrink by orders of magnitude
until its radius is comparable to that of Jupiter.
However, 
planets spin up as they contract as a consequence of angular momentum conservation.
Eventually they hit up against a centrifugal barrier: when the planet spins at breakup velocity,
it cannot contract further without shedding spin angular momentum.
To aggravate the problem, the spin of the planet at the end of runaway growth will already be close to breakup if it accretes mass with the Keplerian shear velocity
across the Hill sphere.

Possible mechanisms to dispose of excess angular momentum include expulsion of material, once breakup is reached, into a circumplanetary disc \citep[CPD;][]{WardCanup2010} and magnetic interaction between the planet and such a disc \citep{TakataStevenson1996}. 
\citet{Batygin2018} demonstrated that magnetic planet--CPD interaction can explain the observed sub-breakup spins of young extra-solar gas giants \citep{Bryan2018}. 
The same spin regulation mechanism can be invoked for isolated brown dwarfs and T-Tauri stars (objects
without a stellar companion); in these cases the CPD and primary disc are one and the same \citep[see][]{Koenigl1991,ArmitageClarke96}.\footnote{Magnetic braking also occurs when spin
angular momentum is carried away by a magnetized wind emanating from the object. 
Wind braking seems too weak to explain the slow rotation of low-mass stars and brown dwarfs that are a few Myr old \citep{Kawaler88,Bouvier2014,Moore2019}.}
While \citet{Batygin2018} focused on the terminal rotation of planets and evolved them from an initial condition of twice the radius of Jupiter, here we expand the scope of the theory to cover earlier times, and ask whether planets can contract starting from as large a radius as the Bondi/Hill radius.
We utilize the \citet{Christensen2009} scaling to relate a planet's magnetic field to its convective luminosity and thence to its KH contraction time, thereby self-consistently evolving the planet's radius and magnetic field over the entire duration of post-runaway accretion.
  
The rest of this letter is organized as follows. 
We describe the planetary magnetic field and its coupling to the CPD in Section \ref{sec:coupling}, and calculate the planet's spin evolution in Section \ref{sec:evolution}. We compare our theory with observations of rotation periods in Section \ref{sec:observations}, and summarize in Section \ref{sec:summary}. 

\section{Magnetic coupling}\label{sec:coupling}

We present a model for the magnetic interaction between a nascent planet and its CPD. The treatment is similar to that of \citet{Batygin2018}. 
We omit order-unity coefficients to focus on scaling relations.

\subsection{Magnetic field}\label{sec:field}

\citet{Christensen2009} proposed that the magnetic field strength $B$ of a planet of mass $M$ and radius $R$ is determined by equipartition of energy, with the magnetic energy density comparable to the kinetic energy density of the convective flow that transports the planet's internal luminosity $L$: \begin{equation}\label{eq:B}
B^2\sim \rho v_{\rm conv}^2\sim \rho^{1/3}\left(\frac{L}{R^2}\right)^{2/3},   
\end{equation}  
where $\rho\sim M/R^3$ is the planet's mean density, $v_{\rm conv}$ is the convective velocity, and $L/R^2\sim \rho v_{\rm conv}^3$ is the convective flux. \citet{Christensen2009} demonstrated that this scaling fits both solar system planets and rapidly rotating stars. Recent observations seem to also validate this scaling for hot Jupiters; these emit a higher internal luminosity compared to Jupiter, and are therefore expected to maintain a stronger field \citep{YadavThorngren2017,Cauley2019}.

Throughout most of the evolution discussed here, planets remain larger than $2 R_{\rm J}$, where $R_{\rm J}$ is the radius of Jupiter \citep{GC_Endgame}. Under such conditions, electron degeneracy and electrostatic interactions are negligible, and the planets can be modelled with an ideal gas equation of state. The effects of degeneracy, which might play a role during the latest stages of contraction, are discussed in the appendix. For non-degenerate objects, the KH time-scale is given by
\begin{equation}\label{eq:t_kh}
t_{\rm KH}\sim\frac{GM^2}{RL},
\end{equation}
where $G$ is the gravitational constant. We rewrite equation \eqref{eq:B} as
\begin{equation}\label{eq:B1}
B\sim \left(\frac{GM^2}{R^4}\right)^{1/2}\left(\frac{P_{\rm break}}{t_{\rm KH}}\right)^{1/3},    
\end{equation}
with $P_{\rm break}\sim (GM/R^3)^{-1/2}$ denoting the planet's breakup rotation period.

\subsection{Truncation radius}\label{sec:truncation}

If the magnetic field is strong enough, the CPD does not extend all the way to the planet's surface. Close enough to the planet, the magnetic energy density exceeds the kinetic energy density of the accretion flow, truncating the disc at an inner radius
\begin{equation}\label{eq:rt}
R_{\rm t}\sim\left(\frac{\mu^4}{GM\dot{M}^2}\right)^{1/7},    
\end{equation}
where $\mu=BR^3$ is the planet's magnetic dipole moment and $\dot{M}$ is the mass accretion rate \citep{ElsnerLamb77,GhoshLamb79,Koenigl1991,OstrikerShu1995,MohantyShu2008}. 
Equation \eqref{eq:rt} is appropriate for spherically symmetric
accretion, or for accretion through an equatorial disc in which
the flow transitions from Keplerian to magnetospheric
over a radial length scale comparable to the disc
radius \citep[e.g.][section 15.2]{ShapiroTeukolsky1983}.
The geometry of our problem is different;
three-dimensional simulations of planets
embedded in circumstellar discs find that a planet accretes mainly through its poles, and forms a decretion disc at the equator \citep{Tanigawa2012,Morbidelli2014,Fung2015,FungChiang2016,Szulagyi2016}. We ignore these complications and assume that equation
\eqref{eq:rt} gives the truncation radius of the equatorial
decretion disc. The same assumption was made by \citet{Batygin2018}. 
If nothing else, the dipole geometry of the 
magnetic field, whose energy density dominates at $r < R_{\rm t}$,
should help to enforce a roughly spherical magnetosphere.

In \citet{GC_Endgame} we explained that during post-runaway accretion the planet adjusts its contraction to satisfy $t_{\rm KH}=M/\dot{M}$.
This condition was derived neglecting spin angular momentum;
however, we will find below that magnetic spin-down is
efficient and that contraction can proceed on the KH time-scale,
with the planet radiating
away the accretion luminosity $L=GM\dot{M}/R$. Using this relation, and substituting $B$ from equation \eqref{eq:B1}, we rewrite equation \eqref{eq:rt} as
\begin{equation}\label{eq:rt1}
\frac{R_{\rm t}}{R}\sim\left(\frac{t_{\rm KH}}{P_{\rm break}}\right)^{2/21}>1.    
\end{equation}
The magnetic field always truncates the disc ($R_{\rm t}>R$) in our case because the thermal time-scale $t_{\rm KH}$ is orders of magnitude longer than $P_{\rm break}$ (see Section \ref{sec:evolution}).

\subsection{Magnetic torque}\label{sec:torque}

Magnetic field lines that originate from the planet rotate at its spin frequency $\omega$ and puncture the CPD. At any given radius $r$ in the disc, if the Keplerian orbital angular velocity $(GM/r^3)^{1/2}$ differs from $\omega$, the field lines are twisted by the disc, generating a counter-torque. The total torque that the disc exerts on the planet is given by 
\begin{equation}\label{eq:torque}
T\sim\int_{R_{\rm t}}{\frac{\mu^2}{r^4}{\rm d}r}\sim B^2R^3\left(\frac{R}{R_{\rm t}}\right)^3
\end{equation}
\citep[for details, see][]{LivioPringle92,ArmitageClarke96,SpaldingBatygin2015,Batygin2018}.
The magnitude and sign of the torque are dictated by the inner edge of the disc at $R_{\rm t}$. 
If the inner edge rotates slower (faster) than the planet, the torque will spin the planet down (up). Following \citet{Batygin2018},
we assume that the CPD is connected to
the larger circumstellar disc such that whatever angular momentum 
(of whatever sign) is transferred from the planet to the CPD 
is subsequently transferred to the nebula at large.
In equilibrium, the planet nearly corotates with the disc's edge: $\omega^2\sim GM/R_{\rm t}^3$.  

\subsection{Electrical conductivity}\label{Sec:conductivity}
We have assumed in the above that the disc is sufficiently
electrically conductive that it couples to the planetary
magnetosphere. A measure of the coupling is the magnetic
Reynolds number $\mathcal{R}_{\rm m}$ which needs to be $\gg 1$ 
(otherwise, field lines diffuse rather than advect and twist under the Keplerian disc flow).
\citet{Batygin2018} evaluated $\mathcal{R}_{\rm m}$ using the Keplerian velocity and the disc scale height (their equation 4).\footnote{This choice yields a value intermediate between other choices for $\mathcal{R}_{\rm m}$: a smaller one which replaces the Keplerian velocity with the disc sound speed (this $\mathcal{R}_{\rm m}$ is typically used to gauge whether the disc can sustain magneto-rotational turbulence; e.g. \citealt{Fleming2000}), and a larger one which replaces the disc scale height with the disc radius (this $\mathcal{R}_{\rm m}$ was used by \citealt{TurnerSano2008} to gauge whether toroidal fields could be generated by disc Keplerian shear from radial fields).} 
We will adopt this same definition, 
and use their result that $\mathcal{R}_{\rm m} > 1$
when the temperature $T > 750\textrm{ K}$, which follows from
thermal ionization of trace alkali metals (their fig.~1). 
This prescription overestimates the electrical conductivity
insofar as charge-adsorbing dust grains are ignored,
but also underestimates the conductivity
because it neglects non-thermal
sources of ionization, e.g. ultraviolet radiation. 
For simplicity we take $T$ to be that of the planet's photosphere. 
Our value of $T$ may be an underestimate
because it ignores
other sources of heating such as shocks or dissipative
magnetospheric processes; these appear implicated
by detections of H$\alpha$ emission
from planet candidates
\citep{Wagner2018,Haffert2019,Thanathibodee2009}.

In Fig.~\ref{fig:mag_panels} we mark, for each of the evolutionary tracks described in Section \ref{sec:evolution}, the time when the accreting planet's photosphere exceeds $750$ K.
The model for
$\kappa= 10^{-1}\textrm{ cm}^2\textrm{ g}^{-1}$ and
$\beta=15$ does not exceed this threshold temperature
and so may not be magnetically coupled, modulo the many
uncertainties regarding heating and ionization listed above.
The other models
(including all the models that reach a high mass)  
are nominally coupled starting at times
$t\gtrsim 0.1$--$1 \textrm{ Myr}$,
depending on $\beta$ (the accretion history) and $\kappa$ (the opacity).
This helps to justify our analysis of the observations in Section \ref{sec:observations},
which are of suitably old 
and massive
systems. 

\section{Spin evolution}\label{sec:evolution}

\begin{figure}
	\includegraphics[width=\columnwidth]{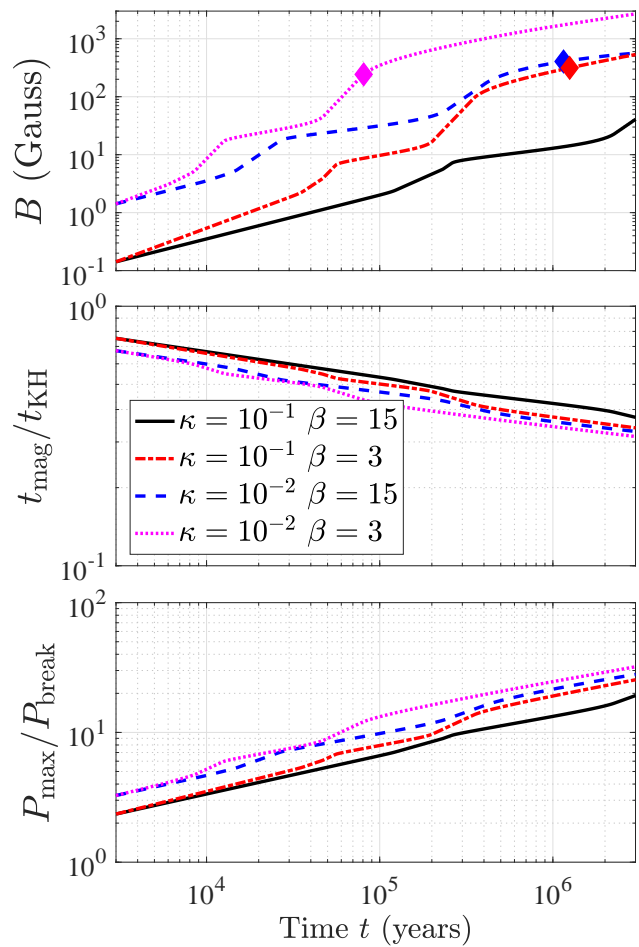}
	\caption{Evolution of planetary spin and magnetic field
          post-runaway, for the four models in \citet{GC_Endgame}.
Time is measured since the end of runaway growth, after which the
planet mass doubles on a time-scale that grows with mass as
$M/\dot{M}\propto M^\beta$.  The opacity $\kappa$ at the 
planet's radiative--convective
boundary 
is given in units of $\textrm{cm}^2\textrm{
  g}^{-1}$. The magnetic field $B$ ({\it top panel}) is given by
equation \eqref{eq:B1}. The magnetic spin-down time $t_{\rm mag}$
({\it middle panel}) is given by equation \eqref{eq:t_mag1} and is
nominally shorter than the Kelvin--Helmholtz contraction time $t_{\rm
  KH}$ by factors of a few. The inequality $t_{\rm mag} \lesssim
t_{\rm KH}$ implies that the rotation period is regulated to be less
than the breakup period $P_{\rm break}$ and that the planet contracts
on the thermal time without
much interference from centrifugal forces. The planet can spin down
until it corotates with the magnetic truncation radius of the
circumplanetary disc, at a period $P_{\rm max}$ ({\it bottom panel}),
given by equation \eqref{eq:p_max}. Diamonds mark when the accreting planet's photosphere becomes 
hot enough to thermally ionize the disc's inner edge to magnetic Reynolds numbers $\mathcal{R}_{\rm m}>1$ and thereby couple it to the planet's magnetic field. This threshold condition
is conservative insofar as other heating and ionization
mechanisms are neglected
(see Section \ref{Sec:conductivity}).
The solid black curve never crosses this threshold.
}
	\label{fig:mag_panels}
\end{figure}

If spin angular momentum were strictly conserved, 
a contracting planet would eventually reach breakup
rotation and cease contracting (Section \ref{sec:introuction}). 
However, the disc can remove
spin angular momentum by magnetic torques.
Shedding as much as the breakup angular momentum
takes a time
\begin{equation}\label{eq:t_mag}
t_{\rm mag}\sim\frac{MR^2\sqrt{GM/R^3}}{T}.
\end{equation}
Using equations \eqref{eq:B1}, \eqref{eq:rt1}, and \eqref{eq:torque}:
\begin{equation}\label{eq:t_mag1}
\frac{t_{\rm mag}}{t_{\rm KH}}\sim\left(\frac{P_{\rm break}}{t_{\rm KH}}\right)^{1/21}\lesssim 1.
\end{equation}
Since $P_{\rm break}$ is many orders of magnitude
smaller than $t_{\rm KH}$, equation \eqref{eq:t_mag1}
implies 
at face value
that $t_{\rm mag} < t_{\rm KH}$, but only by 
a factor of a few because $P_{\rm break}/t_{\rm KH}$ is taken to a weak power.
This result is relatively insensitive to the magnetic field $B$ because of two competing effects that nearly cancel out: a stronger field amplifies the torque exerted at a given radius $r$, but it also truncates the disc at a larger $R_{\rm t}\propto B^{4/7}$. The total torque scales as $T\propto B^2/R_{\rm t}^3\propto B^{2/7}$.

Equation \eqref{eq:t_mag1} indicates that, modulo an uncertain order-unity coefficient, 
magnetic coupling to CPDs may spin planets down somewhat faster than contraction spins them up. We conclude that angular momentum does not present a significant barrier to contraction, which can proceed on approximately the KH time-scale while the planet is regulated to spin slower than breakup. This helps to justify previous one-dimensional calculations that did not consider spin.

In Fig. \ref{fig:mag_panels} we plot four example evolutionary tracks for spin and magnetic field. The tracks represent different post-runaway cooling and accretion models, as computed in figs 1--4 of \citet{GC_Endgame}. 
In that paper, we parametrized the post-runaway accretion rate $\dot{M}$ with a power law $M/\dot{M}\propto M^\beta$, and numerically solved the planet's concurrent accretion and contraction history. 
We accounted for zones of hydrogen dissociation and ionization in the planet's interior, 
and also considered two values for the opacity at the radiative-convective boundary, $\kappa = 10^{-1}$ and $10^{-2}$ $\textrm{cm}^2\textrm{ g}^{-1}$, to account for gas with and without dust, respectively. 
In the $\beta = 15$ models,
which are inspired by planets
opening gaps in practically inviscid discs,
gas accretion drops more drastically after the end of runaway growth, producing planets with a final mass slightly less than Jupiter 
($0.8\,M_{\rm J}$).
By comparison, in the $\beta=3$ models,
which derive from planets opening gaps in more viscous discs,
masses grow to about $5 \,M_{\rm J}$. In low-opacity models,
planets cool and contract to a radius of approximately $2\,R_{\rm J}$ by the time the disc expires at
$t_{\rm disc}=3\textrm{ Myr}$.
In the high-opacity models,
the planets remain inflated
at about $7 \, R_{\rm J}$ 
at that time.

While the four combinations of $\beta$ and $\kappa$ span a range of possible evolutions, Fig. \ref{fig:mag_panels} demonstrates that our conclusions 
for spin appear
robust.
Despite 
a wide range of possible magnetic field strengths,
the magnetic spin-down time $t_{\rm mag}$ is
nominally
always shorter than the KH cooling time $t_{\rm KH}$.
As the planet contracts, its breakup period $P_{\rm break}$ decreases, whereas $t_{\rm KH}$ increases. Thus $P_{\rm break}/t_{\rm KH}$ decreases with time, and by
extension so does $t_{\rm mag}/t_{\rm KH}$
by equation \eqref{eq:t_mag1}.

How much slower does the planet spin relative to breakup?
As discussed in Section \ref{sec:torque}, planets 
seek an equilibrium where they corotate with the disc's magnetic truncation radius $R_{\rm t}$.
Using equation \eqref{eq:rt1}, corotation sets a limiting rotation period given by
\begin{equation}\label{eq:p_max}
\frac{P_{\rm max}}{P_{\rm break}}\sim\left(\frac{R_{\rm t}}{R}\right)^{3/2}\sim\left(\frac{t_{\rm KH}}{P_{\rm break}}\right)^{1/7} 
\end{equation} 
as plotted in the bottom panel of Fig. \ref{fig:mag_panels}.
The time-scale to establish corotation 
is comparable to $t_{\rm mag}$ if
the planet starts at breakup, and shorter
than $t_{\rm mag}$ if it starts slower than
breakup. 
We argued from equation \eqref{eq:t_mag1}
that $t_{\rm mag}/t_{\rm KH} < 1$,
which would imply that planets maintain corotation
as they contract.
However, the margin by which $t_{\rm mag}$
is less than $t_{\rm KH}$ is small and uncertain 
(owing to the weak exponent in equation \ref{eq:t_mag1}
and the unknown coefficient), and so 
we allow for the possibility that corotation equilibrium
might not quite be reached. Thus it is safer to
regard $P_{\rm max}$ as an upper limit on the true
spin period (and $P_{\rm break}$ as a lower limit).
By the end of the disc's lifetime, when $t_{\rm KH}=t_{\rm disc}=3\textrm{ Myr}$ \citep[e.g.][]{GC_Endgame}, planets rotate up to 20--30 times slower than breakup.

\section{Observations}\label{sec:observations}

\citet{Bryan2018} measured the rotation periods of several directly-imaged planets and low-mass
brown dwarfs, compiling a sample that included theirs and previous observations. 
These objects have ages of 2--300 Myr, and their parent discs have all completely dissipated.
We plot the rotation periods and masses of the \citet{Bryan2018} sample in Fig. \ref{fig:obs}, overlaying our theoretical estimates of the maximal rotation period $P_{\rm max}$ relative to the breakup period
$P_{\rm break}$.
Note that $P_{\rm break} \propto R^{3/2}$, and the radius $R$ at the time of disc dispersal depends on atmospheric opacity; the theoretical curves in Fig.~\ref{fig:obs} employ two possible radii,
$2 \, R_{\rm J}$ for dust-free atmospheres, and $7 \, R_{\rm J}$ for dusty ones \citep{GC_Endgame}.

Fig.~\ref{fig:obs} demonstrates that real-life gas giants and brown dwarfs respect the upper bound
on spin periods set by magnetic coupling to CPDs.
If we imagine that planets start at $P/P_{\rm break} \sim P_{\rm max}/P_{\rm break} \approx$ 20--30 at the time
of disc dispersal, and subsequently preserve their spin angular momentum while contracting
and spinning up as $P/P_{\rm break} \propto R^{1/2}$, then the final observed $P/P_{\rm break}$
will drop from its initial value by a factor of a few. This scenario may explain 
why many of the data points appear to cluster around $P/P_{\rm break} \approx 10$.

Whether there is a trend in $P/P_{\rm break}$ with $M$ is unclear. \citet{Bryan2018} did not find any of statistical significance. Fig. \ref{fig:obs} gives the same impression, especially if we omit the one point with the lowest mass---and note that there is an observational bias against detecting slowly rotating, low-mass objects, as these will be among the faintest, with atmospheric lines hardest to spectrally resolve. The absence of a correlation appears consistent with theory; the 1/7 power in equation (\ref{eq:p_max}) flattens all trends. More data are needed at low $M$ to see whether this agreement continues to hold.
  
\section{Summary}\label{sec:summary}

\begin{figure}
	\includegraphics[width=\columnwidth]{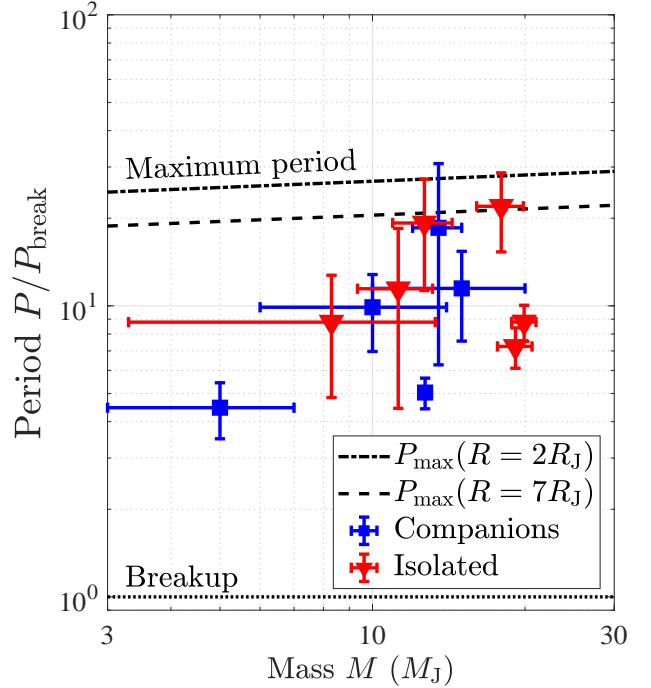}
	\caption{Rotation periods relative to breakup of young
          sub-stellar companions (blue squares) and isolated brown dwarfs (red
          triangles), 
taken from \citet{Bryan2018} with updated error bars
(M. Bryan 2019, personal communication). The sloped dashed
black lines show the longest rotation periods $P_{\rm max}$, normalized
to breakup periods $P_{\rm break}$, to which objects may be
spun down by magnetic coupling to circumplanetary (or circum-brown
dwarf) discs. The ratio $P_{\rm max}/P_{\rm break}$, given by
equation \eqref{eq:p_max} with $t_{\rm KH}$ equated to the disc
lifetime $t_{\rm disc}=3\textrm{ Myr}$, is evaluated for two
values of the object's radius 
$R=2R_{\rm J},7R_{\rm J}$;
these give different
breakup periods at the time of disc dispersal ($P_{\rm break} \propto
R^{3/2}$), with the smaller radius corresponding to contraction of a
relatively low-opacity dust-free atmosphere.
After the disc dissipates, continued contraction at fixed angular
momentum lowers $P/P_{\rm break}\propto R^{1/2}$ by a factor of a few
below the sloped black lines, possibly explaining why the observations
cluster there.}
	\label{fig:obs}
\end{figure}

During the initial cooling-limited phase of its growth, 
a nascent gas giant fills its gravitational sphere of influence
(out to the Bondi or Hill radius) as it accumulates mass
at its periphery faster than it can contract. Accretion
in this phase eventually runs away with the increasing
self-gravity of the gaseous envelope. 
When runaway growth ends---because of 
gaps opened by the planet in its parent disc,
and limitations in the rate at which the disc
can transport mass---the planet's radius
is finally free to begin contracting
on the Kelvin--Helmholtz cooling time-scale $t_{\rm KH}$, 
which is initially as short as a few $10^3$ years.
During this post-runaway phase, 
accretion
continues even as the planet shrinks 
\citep{Mordasini2012,GC_Endgame}.
The radius must decrease by several orders of magnitude
before it attains the present-day observed 
value for Jupiter. 
The problem is that without a mechanism to remove the planet's
spin angular momentum, 
contraction stalls as the planet is inevitably spun up
to breakup speed.

In this letter we considered magnetic coupling of the planet to a circumplanetary disc (CPD) as a means to shed angular momentum and spin the planet down \citep{TakataStevenson1996,Batygin2018}.
New theoretical arguments and observations link the planet's magnetic field to its luminosity  \citep{Christensen2009,YadavThorngren2017,Cauley2019} and thereby to $t_{\rm KH}$. These connections enabled us to consistently compare the magnetic spin-down time $t_{\rm mag}$ to $t_{\rm KH}$ throughout the planet's contraction history, and to extend the calculation of \citet{Batygin2018} to earlier times, when the planet is much larger than present-day Jupiter.
The theory applies also for objects without a stellar host 
(e.g. isolated brown dwarfs); the only requirement is
that the object be surrounded by a disc that can
shuttle angular momentum away to large distance.

We found that at any given time during contraction, $t_{\rm mag}/t_{\rm KH}\sim(P_{\rm break}/t_{\rm KH})^{1/21}$, where $P_{\rm break}$
is the breakup rotation period. Since $P_{\rm break}/t_{\rm KH}\lll 1$,
$t_{\rm mag}/t_{\rm KH} \lesssim 1$, indicating that 
planets are marginally
able to shed their angular momentum while contracting on the Kelvin--Helmholtz time-scale.
That is, the angular momentum barrier to planetary contraction
can be largely removed by CPDs interacting with
primordially strong planetary magnetic fields. An underlying assumption is that the CPD is sufficiently ionized to couple to the planet's magnetosphere; this assumption might fail at early times, especially for low-mass planets.

In addition to justifying the results of
previous studies of giant planet formation
that did not explicitly consider spin,
we also calculated how the rotation periods of gas giants
evolve during post-runaway accretion. 
We found that planets rotate slower than breakup, with a maximal rotation period
given by $P_{\rm max}/P_{\rm break}\sim (t_{\rm KH}/P_{\rm break})^{1/7}$. This ratio gradually increases with time as long as the CPD can
transfer angular momentum from the planet to the nebula at large.
When the parent
disc finally expires at a time 
$t_{\rm disc}\sim t_{\rm KH}\sim 3 \textrm{ Myr}$, 
$P_{\rm max}/P_{\rm break}\approx$ 20--30 in a variety of 
post-runaway models (Fig. \ref{fig:mag_panels}).
Contraction 
at fixed angular momentum 
after the disc vanishes 
may spin planets back up 
to $P_{\rm max}/P_{\rm break}\sim 10$, potentially explaining 
observed rotation periods of young planets and low-mass brown dwarfs (Fig. \ref{fig:obs}).

While we have focused on the magnitude of planetary spin, future work
can examine the direction of the spin vector, i.e. 
the relative orientations of CPDs and planetary
magnetic/spin axes \citep{Lai2011,SpaldingBatygin2015}. 
The CPD might not
necessarily be aligned with the parent circumstellar disc, particularly in turbulent discs. 
Obliquities of sub-stellar objects
are beginning to be measured \citep{Bryan2019}.

\section*{Acknowledgements} 
We thank Marta Bryan for discussions and for providing the \citet{Bryan2018} data. We thank Konstantin Batygin and the anonymous reviewer for comments which improved the letter. SG is supported by the Heising-Simons Foundation through a 51 Pegasi b Fellowship. 
This work benefited from NASA's Nexus for Exoplanet System Science (NExSS) research coordination network sponsored by NASA's Science Mission Directorate
(NNX15AD95G/NEXSS).


\bibliographystyle{mnras}
\input{mag.bbl}



\appendix
\section{Degenerate planets}\label{sec:degenerate}

Electron degeneracy sets a floor on the radius $R_0\approx R_{\rm J}$ below which planets cannot contract \citep[e.g.][]{ZapolskySalpeter69}. We therefore have to correct our scaling laws for the regime $R\lesssim 2 R_0$, or equivalently $\Delta R/R_0\lesssim 1$, where $\Delta R\equiv R-R_0$.
This regime, in which degeneracy can no longer be neglected, is relevant for the late stages of contraction in the case of low opacities 
(see fig. 3 of \citealt{GC_Endgame}). The cooling time in this regime is given by 
\begin{equation}\label{eq:t_kh_deg}
 t_{\rm KH}\sim \frac{GM^2}{R_0L}\frac{\Delta R}{R_0},   
\end{equation}
which replaces equation \eqref{eq:t_kh}. 
In addition, degenerate planets need lose only a fraction
$\sim$$\Delta R/R_0$
of their angular momentum to contract. Taking these two differences into account, equation \eqref{eq:t_mag1} is replaced by
\begin{equation}\label{eq:t_mag_deg}
\frac{t_{\rm mag}}{t_{\rm KH}}\sim\left(\frac{P_{\rm break}}{t_{\rm KH}}\right)^{1/21}\left(\frac{\Delta R}{R_0}\right)^{19/21}< 1.    
\end{equation}
We conclude that for a given cooling time $t_{\rm KH}$ (in practice
equal to the age of the system), degenerate planets spin down faster. However, using equations \eqref{eq:t_kh_deg}, \eqref{eq:B}, and \eqref{eq:rt}, their equilibrium rotation periods are closer to breakup:
\begin{equation}
\frac{P_{\rm max}}{P_{\rm break}}\sim\left(\frac{R_{\rm t}}{R}\right)^{3/2}\sim\left(\frac{t_{\rm KH}}{P_{\rm break}}\right)^{1/7}\left(\frac{\Delta R}{R_0}\right)^{2/7},  
\end{equation}
which replaces equation \eqref{eq:p_max}.


\bsp	
\label{lastpage}
\end{document}